\documentclass[12pt]{iopart}

\usepackage{graphicx}
\usepackage{color}
\begin{document}

\title[Power-law/exponential transport of electromagnetic field]{Power-law/exponential transport of electromagnetic field
in one-dimensional metallic nanoparticle arrays}

\author{Gang Song$^1$ \& Wei Zhang$^{1,2}$}

\address{$^1$ Institute of Applied Physics and Computational Mathematics, P. O. Box 8009(28), Beijing 100088, P. R. China}
\address{$^2$ Beijing Computational Science Research Center, Beijing 100084, P. R. China}
\ead{zhang$\_$wei@iapcm.ac.cn} \vspace{10pt}

\begin{abstract}
Based on the coupled-dipole analysis and finite-difference time-domain
simulation, we have investigated the surface plasmon
propagation in one-dimensional metallic nanoparticle (NP) chains.
Our systematic studies reveal that the interplay between the localized
plasmon excitation and the lattice collective behavior leads to two phases (I and II)
of different electromagnetic (EM) field transport properties. In phase I, the EM field
decays follow the {\it \textbf{power-law}}. In phase II, the EM field
shows the {\it \textbf{exponential decay}} in the short distance regime
and the {\it \textbf{power-law decay}} in the long distance regime.
Moreover, universal power-law exponents have been found in the long
propagation distance. The two different EM field propagation
behaviors can be transformed to each other by tuning the parameters
of the excitation fields and/or those of the NP chains. The EM field
transport mechanisms we have found are very useful in the design of
plasmonic waveguide with both strong field confinement and efficient
field/energy transfer, which has important applications in integrated nanophotonic circuits.
\end{abstract}

%

%
%
%

\section{Introduction}

Extensive studies have been performed on the optical properties and
their applications of metallic nanoparticles (NPs), especially the
noble metal (such as the gold or the silver) NPs. Metallic NP arrays
show interesting collective behaviors (in the
absorption/reflection/transmission etc.)\cite{wood1902,augu2008,haes2004,fisc2016,ma2017,chan2015} due to the combination effect of near field and far field, for example, Wood anomaly in the
reflection/absorption spectra. Metallic NPs arranged in a line can be used as a waveguide for localized surface
plasmon (LSP) propagation
\cite{maie2003,maie2002,pike2013,park2004,croz2007,yang2008,krav2015,step2005,xiao2006,soli2012,soli2013,yin2005,radk2007,will2011,barr2011,fevr2012,quin1998,mark2007,rass2013,rass2014,rass2014a}.
The dispersion relationships of the modes for LSP propagating in the
chain have been analyzed in both theory and experiment
\cite{pike2013,park2004,croz2007,yang2008}. A series of theoretical
methods such as the Mie scattering method were used to describe the
characteristics of LSP propagation in metallic NP chains under
specific conditions
\cite{quin1998,mark2007,rass2013,rass2014,rass2014a}.

There are a lot of pervious works that point out the exponential
decay as the main behavior of the energy transport in metallic
nanoparticle chain in most conditions. Some calculation results
showed that the non-exponential decay appeared in metallic NP chains
\cite{will2011,quin1998}. In experiment, the propagation behavior of
LSPs with the bright mode showed the non-exponential decay in the
chain with close-packed NPs \cite{soli2012}. Though many theoretical
and experimental studies have been performed in the past decades,
the characteristics of LSP propagating in metallic NP arrays are
still unclear. It is very important to clarify the propagation
behavior of LSP in metallic NP chains, which is very useful in the
area of optical communications.

In this paper, taking the silver NP arrays as the sample systems, we
apply the coupled-dipole (CD) analysis and the finite-difference
time-domain (FDTD) simulation to explore the characteristics of LSP
propagating in metallic NP chains. The interplay between the
localized plasmon excitation in the individual NP and the lattice
collective behavior leads to two phases of different electromagnetic
(EM) field transport properties. New propagation
mechanisms/properties including power-law/exponential decay of EM
fields and universal power-law exponents have been found.
\section{Two phases of EM field propagation}

In our system, a series of Ag NPs with the radius $R$ are arranged
in a line with the lattice constant $l$ and the incident field
polarization angle with respect to the direction of the chain is
$\theta$(see the schematic diagram in figure 1(a)). In the
coupled-dipole analysis, only the LSP on the first NP is excited by
incident light, which could be realized by using a local tip to
guide the light onto the first NP in experiment. We first consider
the cases of the incident light polarization perpendicular
to/parallel to the chain.

The transport of the EM field in one-dimensional (1D) NP chains is
mainly determined by the competition between the loss and the
interaction among the NPs, which is the combination effect of the
individual NP plasmonic property and the collective behavior from
the lattice. The plasmonic property of a single NP can be described
by the polarizability $\alpha_s$ as \cite{zou2006}:
\begin{eqnarray}
\alpha_s=\frac{3i}{2k^3}\frac{\mu m^2j_1(m\rho)[\rho
j_1(\rho)]'-\mu_1j_1(\rho)[m\rho j_1(\rho)]'}{\mu m^2j_1(m\rho)[\rho
h_1^1(\rho)]'-\mu_1h_1^1(\rho)[m\rho j_1(\rho)]'}
\end{eqnarray}
where $\mu$ and $\mu_1$ are the magnetic permeabilities of the NP
and the background, $k$ is the wave-vector in the vacuum, $m$ is the
ratio of the index of the metal to that of the background, the
dielectric constant of Ag is obtained from the reference
\cite{palik}, $j_1$ and $h_1^1$ are the spherical Bessel functions
and $\rho=kR$. This expression is suitable to characterize the
optical responses of sphere metal nanoparticles with the radius up
to tens of nanometers \cite{zou2006}, which also agree with the
results calculated by the FDTD method. The collective behavior is
due to the coupling between the NPs at different lattice positions.
In the cases of $l\gg R$, the interaction between the NP at position
$nl$ and the other NP at different lattice position $n'l$ can be
described as dipole coupling and can be written as \cite{mark2007}:
\begin{eqnarray}
G_k(r)=(\frac{k^2}{|r|}+\frac{ik}{|r|^2}-\frac{1}{|r|^3})e^{ik|r|}
\end{eqnarray}
with $r=nl-n'l$ for the incident light polarization perpendicular
to the chain and
\begin{eqnarray}
G_k(r)=(-\frac{2ik}{|r|^2}+\frac{2}{|r|^3})e^{ik|r|}
\end{eqnarray}
for the incident light polarization along the chain. The
coupled-dipole approach works well in the case of $l\geq 3R$
\cite{mark2007,rass2013,rass2014,rass2014a,zou2006}, which we mainly focus on. Here we apply it to illustrate
the propagation characteristic of the silver NP chains. By solving
the coupled-dipole equations for the dipole moments $D_k(nl)$ for
NPs at position $nl$,
\begin{eqnarray}
D_k(nl)=\alpha_s[E_n+\Sigma_{n'}G_k(nl-n'l)D_k(n'l)]
\end{eqnarray}
with the incident field $E_n=E\delta_{n,0}$, we obtain the dipole
moment of each NP in the chain as \cite{mark2007}:
\begin{eqnarray}
D_k(nl)=E\int_{-\pi/l}^{\pi/l}\frac{exp(iqnl)}{1/\alpha_s-S(k,q)}\frac{ldq}{2\pi},
\end{eqnarray}
where $S(k,q)=2\Sigma_{n>0}G_kcos(qnl)$. The analytical structure of
$1/\alpha_s-S(k,q)$ has important impact on the propagation
characteristics. The systems show quite different EM field transport
behaviors depending on whether the resonant condition
$Re[1/\alpha_s-S(k,q)]=0$ can be satisfied. The important role of
the analytical structure of $1/\alpha_s-S(k,q)$ reflects the
interplay between the individual NP plasmonic properties
($\alpha_s$) and the collective behavior ($S(k,q)$).

We firstly consider the cases with the incident light polarization
perpendicular to the chain ($\theta=90^\circ$). Here, we use the
lattice constant $l$=180nm. The chain with NPs of $R$=50nm is
studied to demonstrate the EM field propagating in Ag NP chains for
an incident wavelength $\lambda$=320nm as shown in Figure 1. Here
$D_k$ is normalized as $D_k/E\alpha_s$. The linear relation of
$\ln(|I_D|^{1/2})$ [the local field intensity $I_D \propto |D_k|^2$]
versus the logarithm of the propagation distance ($\ln(x)$) implies
an power-law decay of the EM field for $\lambda$=320nm.

\begin{figure}
  \includegraphics[width=5.00in]{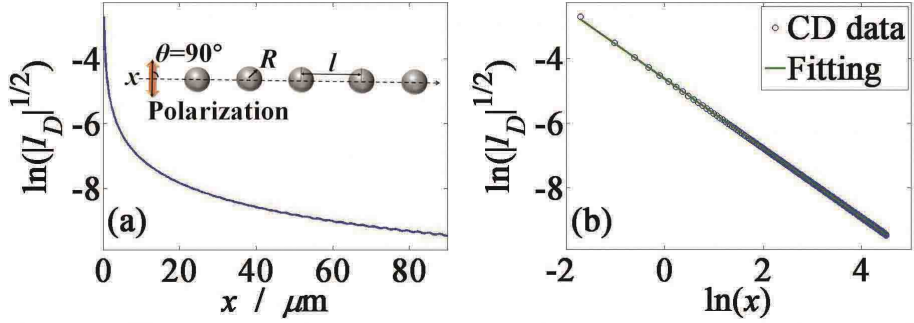}
    \centering\caption{(Color
online) $\ln(|I_D|^{1/2})$ versus the propagation distance $x$ and
$\ln(x)$ for $\lambda$=320nm, respectively. The incident light
polarization is perpendicular to the chain ($\theta=90^\circ$). Here
CD data is the computation results based on coupled-dipole
method.}\label{fig.1}
\end{figure}

\begin{figure}
  \centering\includegraphics[width=3.00in]{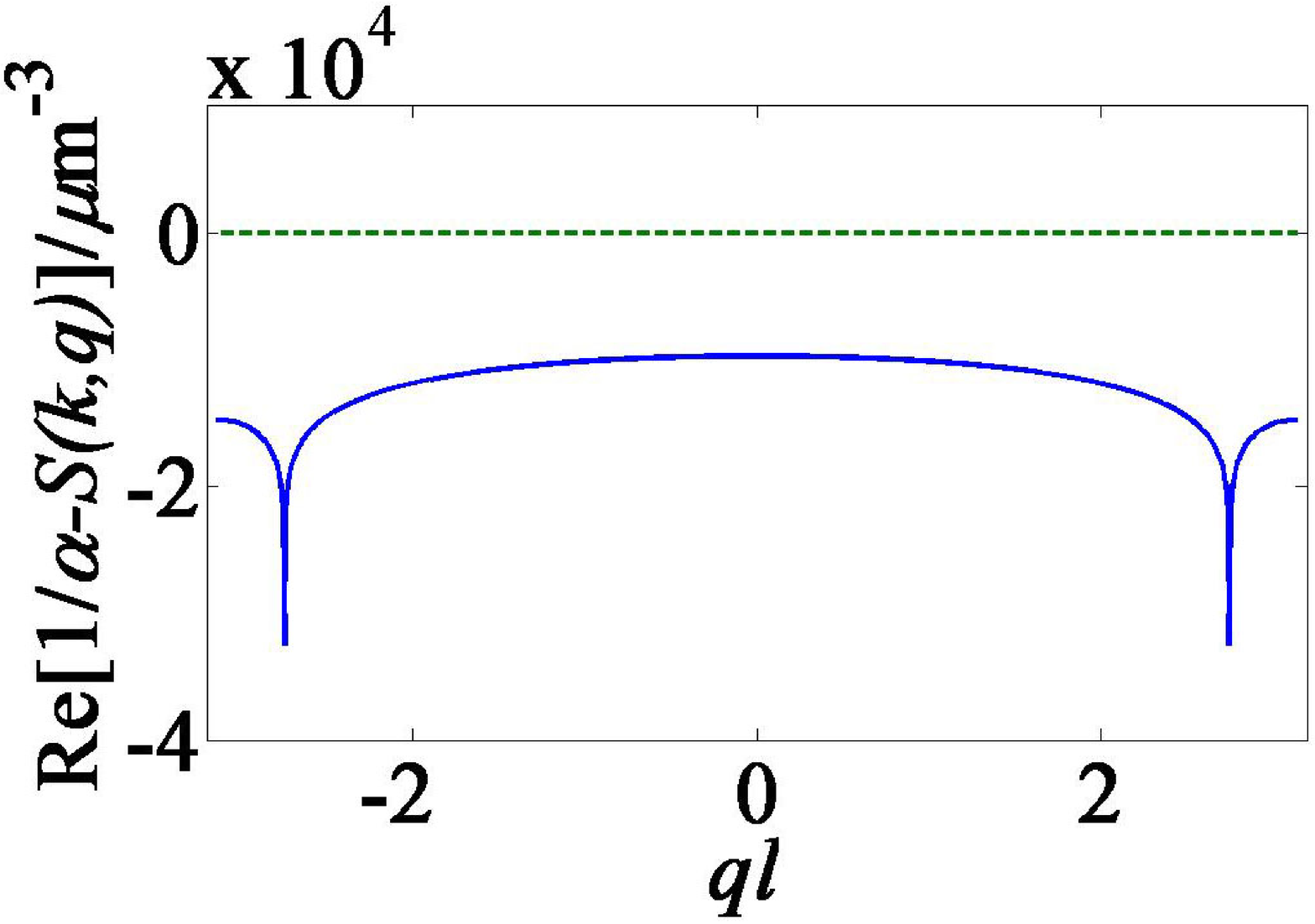}\\
  \caption{(Color
online) $Re[1/\alpha_s-S(k,q)]$ versus $ql$ for $\lambda$=320nm. The
incident light polarization is perpendicular to the chain.}\label{Fig. 2}
\end{figure}

At this wavelength, $Re[1/\alpha_s-S(k,q)]$ is always nonzero as
shown in Figure 2, then the field decay is according to a power law.
At the wavelength $\lambda$=320nm, the absolute value of
$1/\alpha_s$ is larger than that of $S(k,q)$. From Equ.(5) we can
use perturbation theory and obtain
$D_k=\alpha_s^2S_kE=(\alpha_sE)S_k\alpha_s$, where $S_k(nl)=\int
e^{iqnl}S(k,q)ldq/2\pi$. Here we have the physical picture:
$(\alpha_sE)$ describes the dipole moment of a NP with
polarizability $(\alpha_s)$, $S_k$ describes the propagation of the
dipole field, the second $\alpha_s$ describes the coupling with the
other NP. $D_k$ or $S_k$ shows a universal power-law decay behavior,
i.e. $D_k\sim 1/x^\beta$, $\beta$=1 in the long propagation distance
limit. We fit the data from Figure 1(b) and obtain the slope -1.07.
The difference between the fitting power-law exponent 1.07 and
$\beta$=1 is due to the finite fitting propagation distance.

The EM field propagation is determined by the analytical structure
of $1/\alpha_s-S(k,q)$ (as seen from Equ. (5)), which depends on the
wavelength. Then, we take another wavelength to show the other phase
of the EM field propagation. $\lambda$=440nm is adopted as the
working wavelength. $\ln(|I_D|^{1/2})$ versus the propagation
distance $x$ and $\ln(x)$ are shown in Figure 3, respectively. At
$\lambda$=440nm, the linear relation of $\ln(|I_D|^{1/2})$ versus
$x$ indicates the exponential decay in the regime of $x$ from the
beginning to about 4$\mu m$ as shown in the insert of Figure 3(a),
while the linear relation of $\ln(|I_D|^{1/2})$ versus $\ln(x)$
implies the power-law decay in the regime of $\ln(x)$ from 1.81
($x\sim $ 6.12$\mu m$) to the end. Thus, the EM field propagates
following an {\it \textbf{exponential law}} in the short distance
regime and a {\it \textbf{power-law}} in the long distance regime
({\it \textbf{exponential decay}} + {\it \textbf{power-law decay}}).

\begin{figure}
  \centering\includegraphics[width=5.00in]{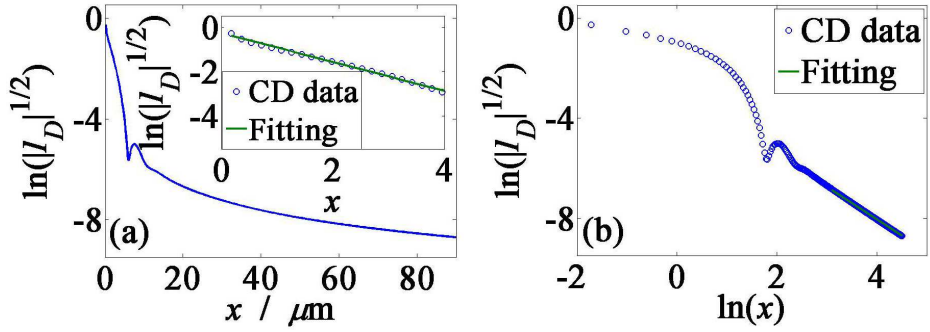}\\
  \caption{(Color
online) $\ln(|I_D|^{1/2})$ versus the propagation distance $x$ and
$\ln(x)$ for $\lambda$=440nm, respectively. The incident light
polarization is perpendicular to the chain.}\label{Fig. 3}
\end{figure}

\begin{figure}
  \centering\includegraphics[width=3.00in]{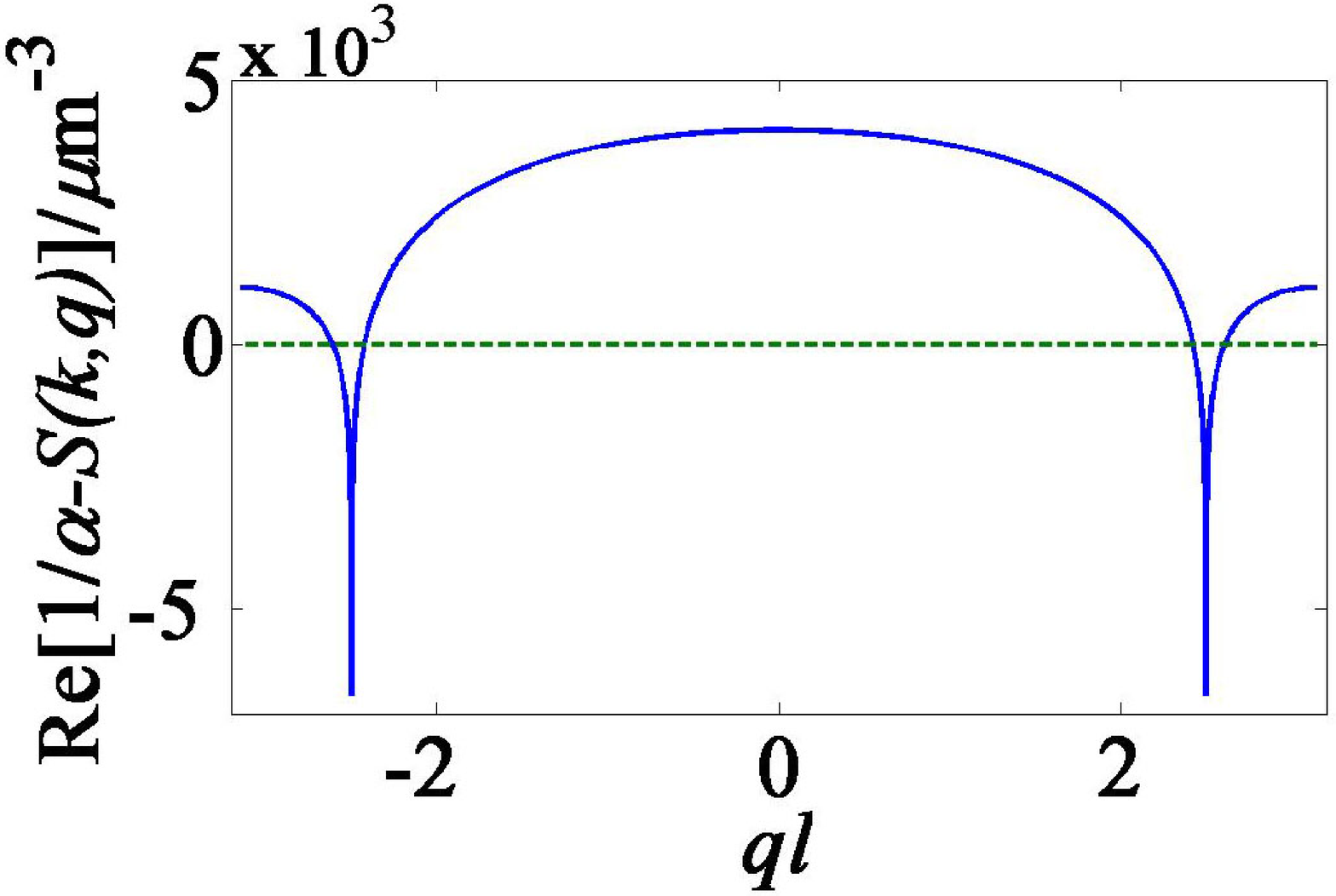}\\
  \caption{(Color
online) $Re[1/\alpha_s-S(k,q)]$ versus $ql$ for $\lambda$=440nm. The
incident light polarization is perpendicular to the chain.}\label{Fig. 4}
\end{figure}

The curve of $Re[1/\alpha_s-S(k,q)]$ versus $ql$ for $\lambda$=440nm
is shown in Figure 4. One can see that there are solutions $q_j$
($j$=1,2...) to the equation $Re[1/\alpha_s-S(k,q)]$=0. As seen from
Figure 4, the solution $q_j$ is around $\pm k$. It is related to the
singularity around $q\pm k$, which is caused by the long-range
coulomb interaction and has the same origin as the Wood anomaly. For
simplicity, we consider a solution $\bar{q}$. The generalization to
the case with multiple solutions is straightforward. Near $\bar{q}$,
$1/\alpha_s-S(k,q)\simeq \frac{\partial Re S(k,q)}{\partial
q}|_{q=\bar{q}}(q-\bar{q})+i \eta \equiv Y$, $\eta=Im
(1/\alpha-S(k,\bar{q}))$. Then, we can rewrite Equ.(5) into two
parts as:
\begin{eqnarray}
D_k(nl)=E\int_{-\pi/l}^{\pi/l}\frac{e^{iqnl}}{Z}\frac{ldq}{2\pi}
=E\int_{-\pi/l}^{\pi/l}\frac{e^{iqnl}}{Y}\frac{ldq}{2\pi}+
E\int_{-\pi/l}^{\pi/l}\frac{e^{iqnl}}{X}\frac{ldq}{2\pi},
\end{eqnarray}
where $Z=1/\alpha_s-S(k,q)$, $1/X=1/Z-1/Y$. The first term on the
right hand of Equ. (6) is in the form of $e^{-nl/L}$
($L=\frac{1}{\eta}|\frac{\partial S(k,q)}{\partial
q}|_{q=\bar{q}}$), showing an exponential decay behavior. While for
the second term, $Re(X)\simeq \frac{2(\partial S/\partial
q)^2}{\partial^2 S/\partial q^2}|_{q=\bar{q}}$ at $q=\bar{q}$.
Considering the very fast change of $Z$ with $q$ near $\bar{q}$ and
the regularization of the singularity of $1/Z$ ($1/Z \rightarrow
1/X=1/Z-1/Y$), the second term shows power-law decay as that in
Figure 1. We can obtain the power-law exponent as 1.29 from fitting
the data in Figure 3(b).

\begin{figure}
  \centering\includegraphics[width=5.00in]{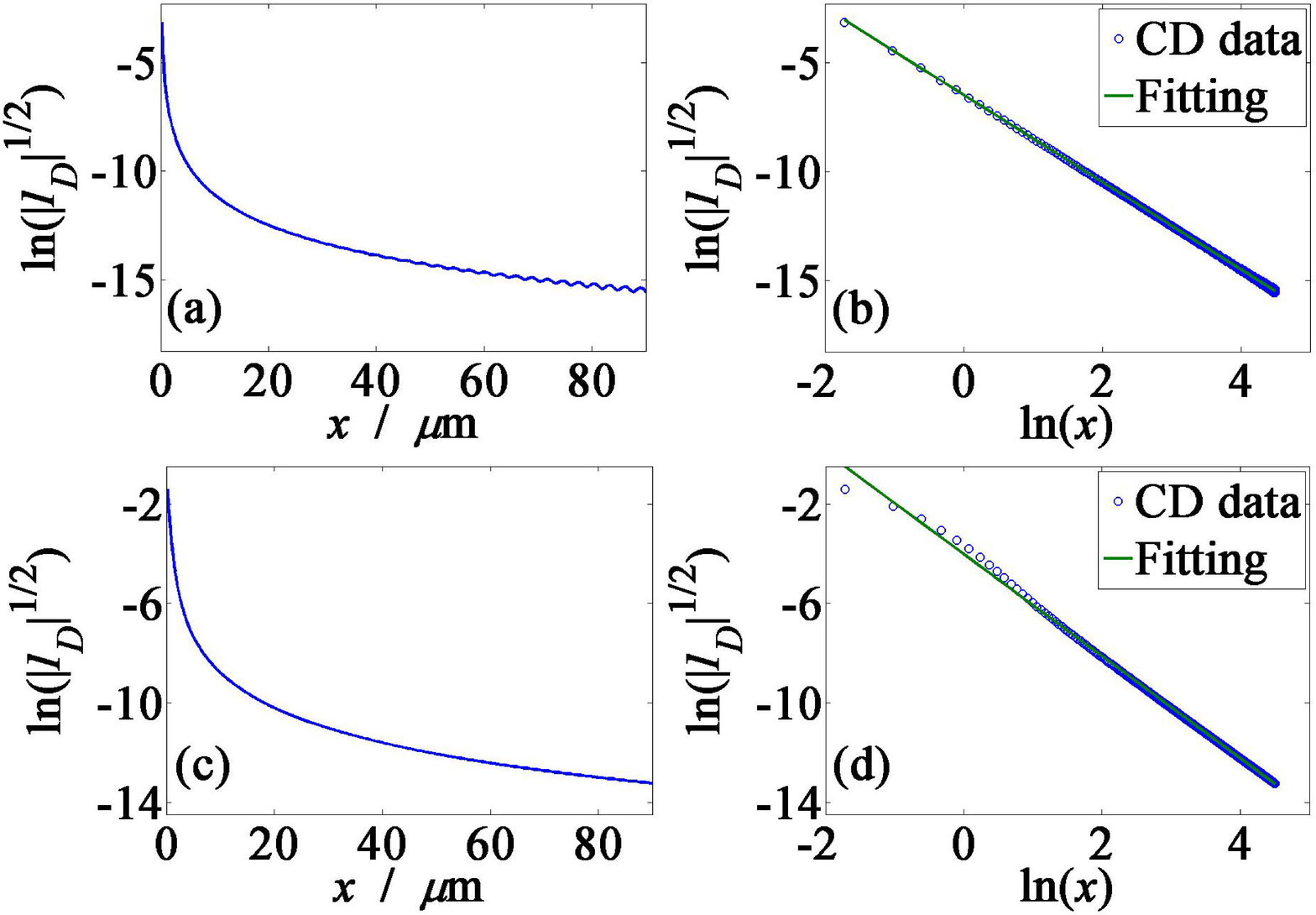}\\
  \caption{(Color
online) $\ln(|I_D|^{1/2})$ versus the propagation distance $x$ and
$\ln(x)$ for $\lambda$=320nm (a-b) and 440nm (c-d), respectively.
The incident light polarization is parallel to the chain.}\label{Fig. 5}
\end{figure}

Then, the polarization of incident light along the chain
($\theta=0^\circ$) is considered. By observing Equ. (2) and Equ.(3),
we can see that the Wood anomaly-like singularity disappears and
expect some changes in the propagation/transport behavior. The
curves of $\ln(|I_D|^{1/2})$ versus the propagation distance $x$ and
$\ln(x)$ for $R$=50nm, $l$=180nm, $\lambda$=320nm and 440nm are
shown in Figure 5, respectively. The clear linear relation of
$\ln(|I_D|^{1/2})$ versus $\ln(x)$ indicates a power-law decay of
the EM fields.

\begin{figure}
  \centering\includegraphics[width=5.00in]{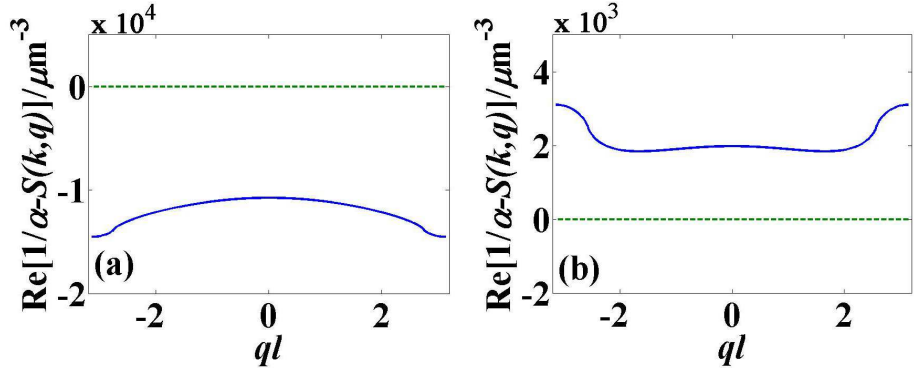}\\
  \caption{(Color
online) $Re[1/\alpha_s-S(k,q)]$ versus $ql$ for $\lambda$=320nm (a)
and 440nm (b), respectively. The incident light polarization is
parallel to the chain.}\label{Fig. 6}
\end{figure}

As shown in Figure 6, there is no solution to the equation
$Re[1/\alpha_s-S(k,q)]$=0 for these two wavelengths $\lambda$=320nm
and 440nm. It is quite different from the case with the polarization
perpendicular to the chain (Figure 2 and Figure 4), which is related
to the disappearance of Wood anomaly-like singularity. Moreover the
power index should also be different. From Equ.(3), the universal
power-law exponent should be 2 in this case. We fit the data from
Figure 5(b) and 5(d) and obtain the slope -2.01 and -2.05 in the
long propagation distance regime, respectively. There may also exist
a small intermediate wavelength range in which there are some $q_0$
to make $Re[1/\alpha_s-S(k,q)]$=0. And the propagation behavior is
the exponential decay + power-law decay.

Here we would like to note that when there is zero of
$Re[1/\alpha_s-S(k,q)]$, the system shows a combination of
exponential decay and power-law decay. In the long distance regime,
the power-law decay dominates. In the short distance regime, it is
also possible that the contribution of the power-law decay term is
larger than that of the exponential decay term. Then the system
shows the effective power-law decay behavior, for example for the
case with $\lambda$=900nm.

To further support our study based on coupled-dipole analysis, we
perform the FDTD simulation of the EM field propagation. In the
simulation, the chain has 200 silver nanoparticles with radius of
50nm and the lattice constant is 180nm. A dipole source is located
at a distance of $l$=180nm to the first nanoparticle along the
chain. The polarization of the dipole source is
perpendicular/parallel to the chain. The detecting point is on the
side of the nanoparticle with 2.5nm away from the interface of the
nanoparticle along the direction of the dipole. The incident
wavelengths are 320nm and 440nm. The gird is 2.5nm. The detecting EM
field intensity is $I_D$. $\ln(|I_D|^{1/2})$ versus $\ln(x)$ ($x$
the propagation distance) is shown in Figure 7.

\begin{figure}
  \centering\includegraphics[width=5.00in]{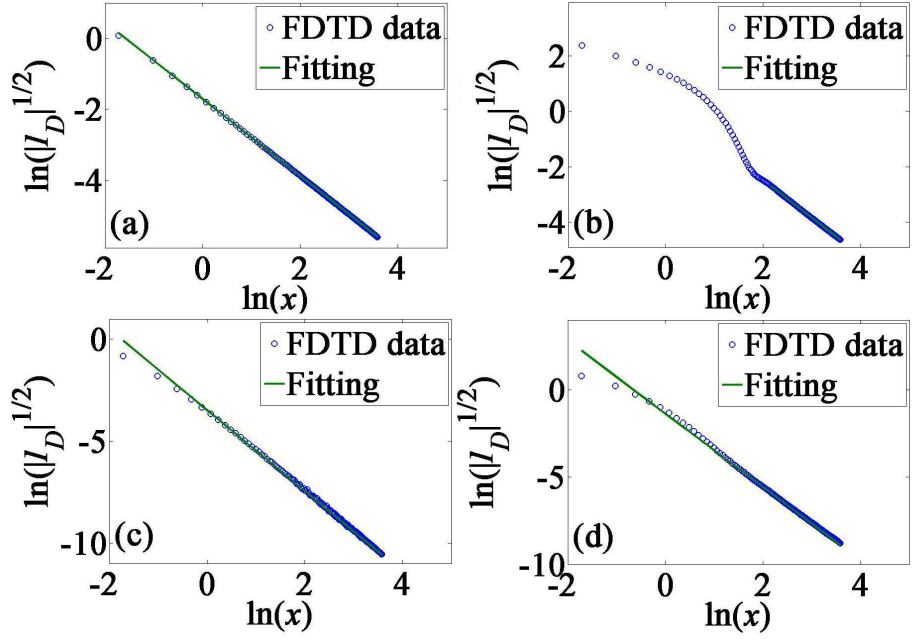}\\
  \caption{(Color
online) $\ln(|I_D|^{1/2})$ versus $\ln(x)$ for the polarization of
the source dipole perpendicular to (a, b) /parallel to (c, d) the
chain with $\lambda$=320nm (a, c) and 440nm (b, d), respectively.
$I_D$ is the EM field intensity, and $x$ is the propagation
distance.}\label{Fig. 7}
\end{figure}

From Figure 7(a) and 7(b), we find two different field transport
characteristics for the source dipole perpendicular to the chain.
The linear relation of $\ln(|I_D|^{1/2})$ versus $\ln(x)$ in Figure
7(a) indicates a power-law decay of EM field with the slope of -1.08
for $\lambda$=320nm. Figure 3 shows an exponential decay+power law
decay of the EM field for $\lambda$=440nm based on CD method. Here,
as shown in Figure 7(b), the linear relation of $\ln(|I_D|^{1/2})$
versus $\ln(x)$ is in the regime from about 1.81($x \sim$ 6.12$\mu
m$) to the end, which implies an {\it \textbf{exponential decay}} +
{\it \textbf{power-law decay}} of EM field with the slope of -1.35
for $\lambda$=440nm. The derivation of the exponent from 1 is due to
the finite fitting regime. From Figure 7(c) and 7(d), the linear
relations of $\ln(|I_D|^{1/2})$ versus $\ln(x)$ for the the
polarization of the source dipole along the chain indicate the
power-law decay of EM field with the slopes of -1.99 and -2.09,
respectively. The power-law exponents based on the couple dipole
calculations and the FDTD simulations are listed in Table 1. As
shown in Table 1, EM field transport characteristics and the
power-law exponents from the FDTD simulations agree well with those
based on coupled-dipole analysis.

\begin{table}
\centering \caption{The  power-law exponents from the couple dipole
(CD) calculations and FDTD simulations.}
\begin{tabular}{lll}
\hline\noalign{\smallskip}
& CD& FDTD\\
\noalign{\smallskip}\hline\noalign{\smallskip}
$\lambda$=320nm, $\bot$ & -1.07 & -1.08 \\
$\lambda$=440nm, $\bot$ & -1.29 & -1.35\\
$\lambda$=320nm, $\|$ & -2.01 & -1.99 \\
$\lambda$=440nm, $\|$ & -2.05 & -2.09\\
\noalign{\smallskip}\hline
\end{tabular}
\end{table}

\section{Universality and modulation of EM field propagation}

We have shown that there are two phases of the EM field propagation
in 1D metallic NP chains, which are determined by the analytical
structure of $1/\alpha_s-S(k,q)$. The parameters of the excitation
field and the structure of the NP chains have important impacts on
the EM field transport behavior. As discussed above, the two phases
of the EM field propagation can be transformed to each other by
tuning the wavelength of the excitation field.

We further explore the EM field propagation with excitation fields
of different polarizations. The calculation results in Figure 8 show
that the EM field propagates according to power-law for
$\lambda$=320nm. The power-law exponents is 1.08  for incident field
polarization angles  $30^\circ$, $45^\circ$, $60^\circ$, $90^\circ$
with respect to the direction of the chain. The power-law exponents
is 1.99 for the polarization angle $0^\circ$. In the presence of
excitation fields  with wavelength $\lambda$=440nm and polarization
angles $30^\circ$, $45^\circ$, $60^\circ$, $90^\circ$, the EM field
propagates following exponential law in the short distance regime
and power-law in the long-distance regime. The power-law exponents
are 1.43, 1.41, 1.40, 1.35 for polarization angles $30^\circ$,
$45^\circ$, $60^\circ$, $90^\circ$. These power-law exponents
obtained by fitting the $\ln(|I_D|^{1/2})-\ln(x)$ curves in the
regimes of finite propagation distance show the combination
power-law decay of EM field with exponents 1 and 2. In the very long
propagation distance regime, the leading contribution is the
power-law decay with exponent 1. For the case of incident field
polarization angle $0^\circ$, the EM field propagates following the
power law with exponent 2.09 (by fitting the calculation data in
finite propagation distance).

The geometric parameters of the NP chain also play an important role
in the EM field transport. The calculation results of EM field
propagation in metallic chains with Ag NPs of radius $R$=40nm, 45nm,
55nm are shown in Figure 9. The general power-law decay of EM field
has been found for the cases with the wavelength $\lambda$=320nm
(incident field polarization parallel to/ perpendicular to the
chain) and the wavelength $\lambda$=440nm (the incident field
polarization parallel to the chain) as shown in Figure 9(a), 9(b)
and 9(c). While for $\lambda$=440nm (incident field polarization
perpendicular to the chain), the EM field transport behavior is
exponential decay + power-law decay. For NPs with different sizes,
the contributions from the exponential decay and the power-law decay
are different, which leads to the change of the transition point
(from the exponential decay to the power-law decay).

\begin{figure}
  \centering\includegraphics[width=5.00in]{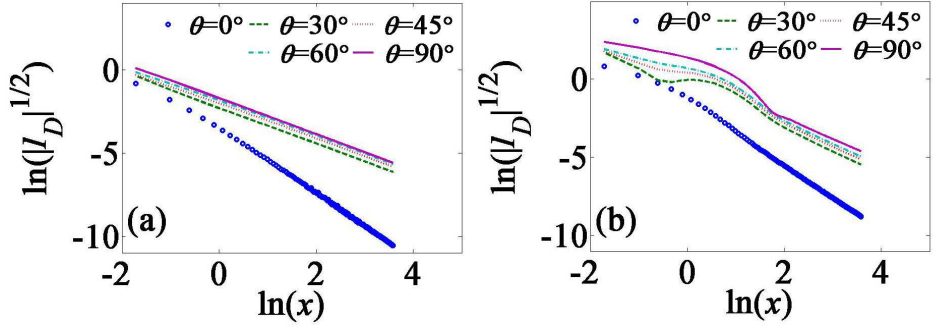}\\
  \caption{((Color
online) $\ln(|I_D|^{1/2})$ versus $\ln(x)$ for the polarization
angles $0^\circ$, $30^\circ$, $45^\circ$, $60^\circ$, $90^\circ$ of
the source dipole with respect to the direction of the Ag chain. (a)
$\lambda$=320nm and (b) $\lambda$=440nm.}\label{Fig. 8}
\end{figure}

\begin{figure}
  \centering\includegraphics[width=5.00in]{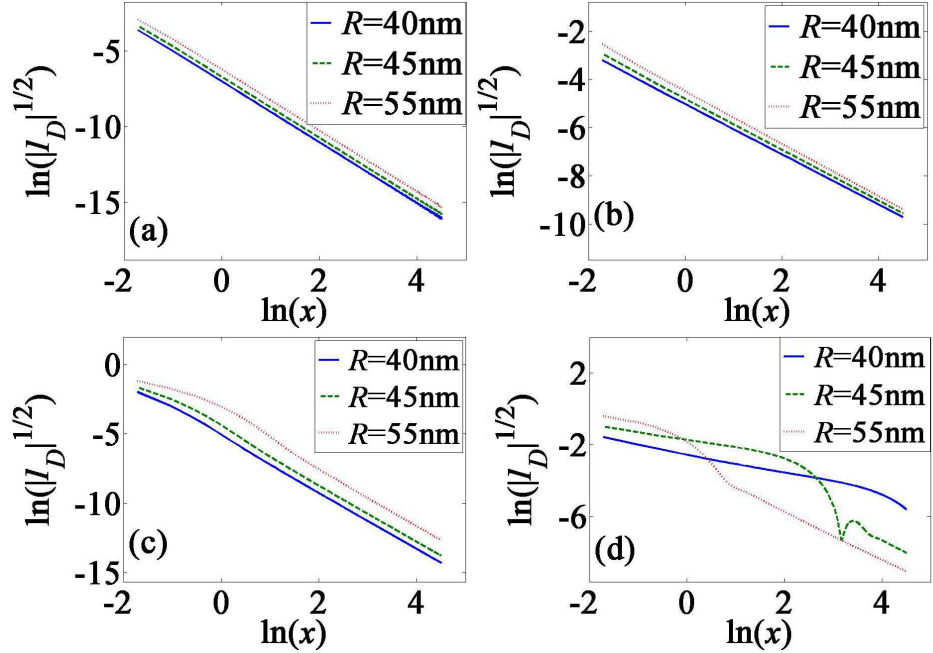}\\
  \caption{(Color
online) $\ln(|I_D|^{1/2})$ versus $\ln(x)$ for Ag chains with
particles of radius $R$=40nm, 45nm,55nm, respectively. The
polarization of the source dipole is parallel to (a, c)
/perpendicular to (b, d) the chain with $\lambda$=320nm (a, b) and
440nm (c, d), respectively.}\label{Fig. 9}
\end{figure}

\begin{figure}
  \centering\includegraphics[width=5.00in]{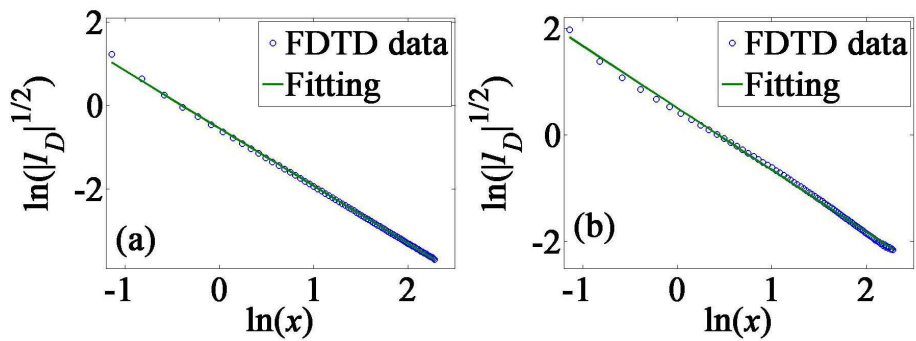}\\
  \caption{(Color
online) $\ln(|I_D|^{1/2})$ versus $\ln(x)$ for the polarization of
the source dipole perpendicular to Ag chain for $\lambda$=300nm (a)
and Au chain for $\lambda$=514nm (b), respectively. $I_D$ is the EM
field intensity, and $x$ is the propagation distance.}\label{Fig. 10}
\end{figure}

We have focused our studies on the systems with large lattice
constant ($l>3R$), in which the CD method works well as supported by
the FDTD simulations. By using  CD method, we are able to provide
analytical solutions and clear physical picture/mechanism, also
gives the universal power-law exponents. We then consider the cases
of lattice constant smaller than $3R$. We use FDTD method to study
the propagation behavior of NP chains with $l<3R$. In the
calculation we use the NP chain with $R$=50nm, $l$=120nm. Silver and
Gold NPs are all considered to demonstrate the universality of the
propagation property. The dielectric constants are obtained from the
reference \cite{palik}. The incident field polarization is
perpendicular to the chain and the wavelength $\lambda$ is 300nm for
Ag NP chain and 514nm for Au NP chain. $\ln(|I_D|^{1/2})$ versus
$\ln(x)$ are shown in Figure 10. As shown in Figure 10, both the two
lines show the linear relation with $\ln(x)$. We fit the two lines
and the two slopes are -1.38 and -1.16. These results point that the
power-law decay may also appear in different metallic NP chains with
different lattice constants.

\begin{figure}
  \centering\includegraphics[width=5.00in]{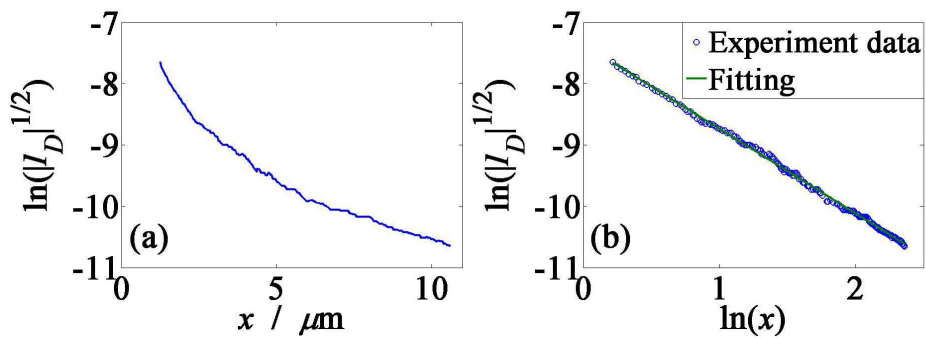}\\
  \caption{(Color
online) (a) $\ln(|I_D|^{1/2})$ versus the propogation distance $x$
based on the the experiment data at $\lambda$=514nm \cite{soli2012};
(b) $\ln(|I_D|^{1/2})$ versus $\ln(x)$ for $\lambda$=514nm.}\label{Fig. 11}
\end{figure}

The non-exponential decay is also pointed out in the experiment
 with close-packed NP chain \cite{soli2012}. The experiment data can
be replotted in the double logarithm form as shown in Figure 11. The
linear relation between $\ln(|I_D|^{1/2})$ versus $\ln(x)$ ($x$ the
propagation distance) indicate a power-law decay of EM field. The
pow-law exponent is 1.38. Hence, our analytical and simulation
methods provide new ways to clarify the propagation behaviors in 1D
metallic NP chains in previous studies
\cite{soli2012,will2011,quin1998}.

\section{Conclusions}

In summary, we have studied the characteristics of the LSP
propagation in one-dimensional metallic nanoparticle chain based on
the coupled-dipole calculations and FDTD simulations. The LSP
propagation shows  different behaviors in different phases,
depending on the analytical structure of $1/\alpha_s-S(k,q)$=0,
which is determined by both the individual NP plasmonic property and
the lattice collective effect. In phase I, a power-law decay of EM
field has been identified. While in phase II, EM field propagates in
the form of  the exponential decay+power law decay, i.e.,
exponential decay in the short distance regime and  power-law decay
in the long distance regime. The universal power-law exponents have
been found for EM field propagation  in the long propagation
distance regime. The results of FDTD simulations agree well with our
theoretical predictions based on coupled-dipole method. It is found
that the incident field (wavelength and polarization) and the
geometric parameters of the NP chains have important impacts on the
EM field transport properties. Our studies reveal a general picture
of the EM field transport in nanoparticle chains, which have
applications in the plasmonic waveguide and integrated nanophotonic
circuits.

\section{Acknowledgement}
This work was partially supported by National Key Research and
Development Program of China (Grant No. 2017YFA0303400), National
Natural Science Foundation of China (Grant Nos. 11774036, 11374039).

\end{document}